\input harvmac.tex
\noblackbox

%
%

%
\ifx\epsfbox\UnDeFiNeD\message{(NO epsf.tex, FIGURES WILL BE
IGNORED)}
\def\figin#1{\vskip2in}
\else\message{(FIGURES WILL BE INCLUDED)}\def\figin#1{#1}\fi
\def\ifig#1#2#3{\xdef#1{fig.~\the\figno}
\goodbreak\midinsert\figin{\centerline{#3}}%
\smallskip\centerline{\vbox{\baselineskip12pt
\advance\hsize by -1truein\noindent\footnotefont{\bf
Fig.~\the\figno:} #2}}
\bigskip\endinsert\global\advance\figno by1}
\def\kk{Kaluza-Klein}
\def\real{I\negthinspace R}

%
%
\lref\kkmon{R.Sorkin, Phys.\ Rev.\ Lett.\ {\bf 51} (1983) 87 \semi
D.Gross and M.Perry, Nucl.\ Phys.\ {\bf B226} (1983) 29.}
\lref\sennew{A.Sen, ``Kaluza-Klein Dyons in String Theory,''
hep-th/9705212.}
\lref\imamura{Y. Imamura, ``Born-Infeld Action and Chern-Simons
Term from Kaluza-Klein Monopole in M-theory,'' hep-th/9706144.}
\lref\hmon{R.Rohm and E.Witten, Ann.\ Phys.\ {\bf 170} (1986) 454
\semi
T.Banks, M.Dine, H.Dijkstra and W.Fischler,
Phys.\ Lett.\ {\bf B212} (1988) 45 \semi
A.Strominger, Nucl.\ Phys.\ {\bf B343} (1990) 167;
E: Nucl.\ Phys.\ {\bf B353} (1991) 565 \semi
R. Khuri, Phys. Lett. {\bf B294} (1992) 325; hep-th/9205051 \semi
J.P.Gauntlett, J.A.Harvey and J.Liu,
Nucl.\ Phys.\ {\bf B409} (1993) 363; hep-th/9211056.}
\lref\mahsch{J.Maharana and J.Schwarz, Nucl.\ Phys.\ {\bf B390} (1993) 3;
hep-th/9207016.}
\lref\senred{A.Sen. Nucl.\ Phys.\ {\bf B404} (1993) 109; hep-th/9207053.}
\lref\gibrub{G.Gibbons and P.Ruback, Phys.\ Lett.\ {\bf B215} (1988) 653.}
\lref\tref{A.Giveon, M.Porrati and E.Rabinovici,
Phys.\ Rep.\ {\bf 244} (1994) 77.}
\lref\ch{C.G.Callan Jr.\ and J.A.Harvey, Nucl.\ Phys.\ {\bf B250}
(1985) 427.}
\lref\ogva{H. Ooguri, C. Vafa, Nucl.Phys. {\bf B463} (1996) 55,
hep-th/9511164; D. Kutasov, Phys. Lett {\bf B383} (1996) 48,
hep-th/9512145; H. Ooguri and C. Vafa, hep-th/9702180.}
\lref\Anselmi{D. Anselmi, M. Bill\'o, P. Fr\`e, L. Girardello,
A. Zaffaroni, Int.J.Mod.Phys. {\bf A9} (1994) 3007, hep-th/9304135.}
%
%

\Title{
\vbox{\baselineskip12pt\hbox{DTP/97/31}\hbox{EFI-97-26}
\hbox{YCTP-P15-97}\hbox{hep-th/9708086}}}
{\vbox{
\centerline{Unwinding Strings and T-duality}
\centerline{of Kaluza-Klein and H-Monopoles} }}
\medskip
\centerline{Ruth Gregory}
\centerline{\it Centre for Particle Theory, University of Durham}
\centerline{\it South Road, Durham, DH1 3LE, U.K.}
\medskip
\centerline{Jeffrey A. Harvey}
\centerline{\it Enrico Fermi Institute and Department of Physics}
\centerline{\it University of Chicago}
\centerline{\it 5640 Ellis Avenue,Chicago, Illinois 60637}
\medskip
\centerline{Gregory Moore}
\centerline{\it Department of Physics, Yale University}
\centerline{\it New Haven, CT 06520}
\bigskip
\bigskip

\centerline{\bf Abstract}
A fundamental string with non-zero winding number can unwind
in the presence of  a Kaluza-Klein monopole. We use this
fact to deduce the presence of a zero mode for the Kaluza-Klein
monopole corresponding to excitations carrying H-electric charge
and we study the coupling of this zero mode to fundamental
strings. We also a describe a T-dual process in which the momentum
of a fundamental string ``unwinds'' in the presence of an
H-monopole. We use the coupling of string winding modes to the
dyon collective coordinate of the Kaluza-Klein monopole to argue that there
are stringy corrections to the Kaluza-Klein monopole which are
in accordance with T-duality.
\medskip
\Date{August 15, 1997}
%
%

\newsec{Introduction}

Supersymmetric string theory compactified on $T^6$ contains
Kaluza-Klein monopoles \kkmon\  carrying charge under
a $U(1)$ arising from Kaluza-Klein reduction of the ten-dimensional
metric, and H-monopoles \hmon\ which carry charge under a
$U(1)$  arising from reduction of the
ten-dimensional Kalb-Ramond field $B_{MN}$.

Consider a \kk\ monopole associated with one of the compact
directions, say $x^5$.
At spatial infinity the  spacetime has
topology $S^3 \times T^5$.  The factor
$T^5$ is a compactifying torus. The factor  $S^3$
has the following interpretation:  $S^3$ is
topologically the Hopf fibration,   an
$S^1$ fibration over
$S^2$.  In the \kk\ monopole $S^3$ is metrically
a squashed three-sphere: at large spatial distances
the radius of the Hopf fiber, parametrized locally
by $x^5$, becomes constant and serves as a
circle for Kaluza-Klein reduction. The base space
$S^2$ is a metrically round $S^2$ and plays the role of  the ``spatial'' points
at fixed radius in \real$^3$.
  Consider now a string
wound around the Hopf fiber.
Although squashed, the
three-sphere is nevertheless a nontrivial $S^1$ fiber
bundle over $S^2$, and in particular the total
space, which is topologically $S^3$, has
$\pi_1 =0$. Therefore, a string wound
around a Hopf fiber at a large distance from the
monopole can be  unwound by motion that keeps it
arbitrarily far away from the monopole.
On the other hand,
the winding number of the string is  a conserved
charge, the H-electric charge, and is coupled to a gauge field $B_{\mu 5}$.
Since this charge is conserved and the charge of the string
changes when the string unwinds, there must be another source
of H-electric charge. The only plausible possibility is that
the \kk\ monopole can also carry H-electric charge and that
the H-electric charge of the string flows onto the monopole
in the unwinding process. We will show below that this is
in fact the case and identify the zero mode of the \kk\ monopole
which carries this charge.

We can also consider a T-dual version of this process. T-duality
takes the \kk\ monopole to an H-monopole and a string with
non-zero winding to a string with non-zero momentum along
the $S^1$. It must then be possible to ``unwind'' a string with
momentum in the presence of an H-monopole, again without bringing
the string near the monopole. This implies that H-monopoles
can carry \kk\ electric charge, and we again identify the relevant
zero mode.

Finally, we use the fact that strings can unwind in a \kk\
monopole background to show that there are stringy corrections
to the \kk\ monopole solution and we argue that these resolve
a number of puzzles regarding T-duality of H-monopoles and
\kk\ monopoles. The interplay between T-duality, 5branes,
and ADE singularities has also been discussed in
 \ogva\Anselmi.

\newsec{Unwinding strings in a \kk\ monopole background}

We begin by considering  Type II string theory compactified
on $T^6$. Many of our considerations also apply to heterotic
strings, but there are some additional subtleties which will
be discussed elsewhere.
Type II theory has a \kk\ monopole solution \kkmon\ with the string metric
given by
\eqn\kmon{ds^2 = dt^2 - U(dx^5 + {\half} R(1-\cos \theta) d \phi)^2 -
U^{-1} (dr^2 + r^2 d \Omega_2^2) }
with
\eqn\udef{U(r) = \left( 1 + {R \over 2r} \right)^{-1} }
and with a flat metric on the remaining five compact coordinates.
The metric is non-singular provided that $x^5$ is periodic with
period $2 \pi R$. In what follows it will be useful to have two coordinate
patches to describe this solution. Note that as $\theta\to0$ we can write
the above metric in a non-singular fashion, but as $\theta\to \pi$, the
metric appears singular. This is simply a coordinate singularity and can
be removed by setting
\eqn\bardef{
{\bar x}^5 = x^5 + R\phi .
}
Since $\phi$ has periodicity $2\pi$, this is clearly a well defined
transition function. The resulting $t,r=$constant surfaces are described
by the Hopf fibration of $S^1$ over $S^2$ to give the total
space $S^3$. Thus although {\it locally} for large $r$ the spacetime
has the appearance of being \real$^4 \times S^1 \times T^5$, it
in fact has global topology \real$^5 \times T^5$.

Now consider a string which is wound once around the $x^5$ direction
and is located at radial coordinate $r_0$ and at the North pole
on the spatial $S^2$. Since $\Pi_1($\real$^5)$ is trivial, any closed
loop is contractable and we should be able to unwind the string. One
way this can happen is for the wound string to be pulled into the
core of the monopole where it then unwinds. From \kmon\ we see
that the proper distance around $x^5$ decreases as we move in towards
the center of the monopole. The wound string can thus decrease its
winding energy by moving in towards the monopole.
Another way of seeing this is to note that in the dimensional reduction
of \kmon\ to four dimensions, the wound string which now appears to
be a particle, picks up a factor of $\sqrt{U}$ in the usual
action for a relativistic test particle. This introduces a modification
to the geodesic equation
%
%
which can be interpreted as a central attractive force.
The string is therefore attracted to the core of the \kk\ monopole
where it can unwind. This process was considered by Gibbons and
Ruback for Nambu strings \gibrub\ and will play an important role in our
analysis of T-duality in section 4.

However, since the spatial topology at infinity is $S^3$
it must also be possible to unwind this string while keeping it arbitrarily
far from the monopole. This is accomplished by the following
trajectory labelled
by   world-sheet coordinates $\sigma, \tau $
with $0 \le \sigma, \tau  \le 1$:

\eqn\unwinda{\eqalign{
& x^5= 2 \pi R \sigma \cr
& \theta = \pi \tau \cr
& \phi = - 2 \pi \sigma \cr
& r=r_0. \cr
 }}
As $\tau \to 1$, $\theta\to \pi$, and we must use the barred
coordinate patch in which ${\bar x}^5=0$ and the string is unwound.
To understand where the  winding charge has gone we need
to analyze the problem in somewhat more detail.

The only fields relevant to the resolution of this puzzle can be
packaged in terms of the dimensional reduction of a five-dimensional
theory containing gravity and an antisymmetric tensor field.  We also add
a scalar field  representing the dilaton.  We thus
for the time being ignore the five additional coordinates needed to
embed the solution into string theory as well as the additional
moduli of string theory.

Therefore, we consider the dimensional reduction of the action
describing the above five-dimensional fields and their coupling to
the string:
\eqn\action{
\eqalign{
S_5 = &{1 \over 2 \kappa_5^2} \int d^5x \sqrt{\cal G} e^{- 2 \Phi}
[ -{\cal R} - 4 (\partial_a \Phi)^2
+ {1 \over 12} {\cal H}_{abc}^2]\cr
&-{1\over4\pi\alpha'} \int d^5x \int d^2\sigma
\delta^{(5)} \left (x^a - X^a(\sigma^A )\right )
\left \{ \sqrt{\gamma}
\gamma^{AB} {\partial X^a\over\partial\sigma^A}
{\partial X^b\over\partial\sigma^B} {\cal G}_{ab} +
\epsilon^{AB} {\partial X^a\over\partial\sigma^A}
{\partial X^b\over\partial\sigma^B} {\cal B}_{ab} \right \}\cr
}}
Our convention is that roman indices run over 5 dimensions, greek from 0 to 3,
capitals
from 0 to 1.
 The signature is mostly minus, and the calligraphic symbols
denote five-dimensional objects which will ultimately be reduced to 4
dimensions. Note that the \kk\ monopole is still a solution within
this theory with a constant dilaton and zero ${\cal B}$-field.
We now dimensionally reduce the action \action\  following \refs{\mahsch,
\senred} setting
\eqn\metric{
ds^2 = -e^{-4 \sigma} [dx^5+A_\mu^1 dx^\mu]^2 + g_{\mu\nu} dx^\mu dx^\nu\ ,
}
\eqn\w{
A_\mu^2 = {\cal B}_{\mu 5}\ ,
}
and
\eqn\anti{
B_{\mu\nu} = {\cal B}_{\mu\nu} + {1\over 2} (A_\mu^1 A_\nu^2 -
A_\nu^1 A_\mu^2) .
}
This theory has an $O(1,1)$ symmetry which we can make manifest by
introducing the two by two matrices
\eqn\makman{L = \pmatrix{0 & 1 \cr 1 & 0 \cr }, \qquad
M = \pmatrix{e^{4 \sigma} & 0 \cr 0 & e^{-4 \sigma} \cr }. }
The field strengths are then defined by
\eqn\eff{
F_{\mu \nu}^{\hat a} = \partial_\mu A_\nu^{\hat a} -
\partial_\nu A_\mu^{\hat a}}
with ${\hat a}=1,2$ and
\eqn\fourax{
H_{\mu\nu\lambda} = \partial_\mu B_{\nu\lambda} - {1\over2}
A_\mu^{\hat a}L_{{\hat a} {\hat b}} F_{\nu \lambda}^{\hat b} + {\rm cyclic}.
}
Note that $B_{\mu \nu}$ now transforms under gauge transformations
of $A$. In particular, under a gauge transformation
$\delta A^1 = d \Lambda^1$, $B$ transforms as $\delta B = \half d \Lambda^1
\wedge A^2$. This means in particular that in the \kk\ monopole background
both $B$ and $A^1$ will be patch dependent with the relation between
the fields in the two coordinate patches being given by
\eqn\patchrules{
\eqalign{
{\bar A^1} &= A^1 - Rd\phi \cr
{\bar B} &= B + {\half} R d\phi \wedge A^2 \cr
}
}

Continuing with the dimensional reduction, we see that the first part of the
action becomes
\eqn\four{
S_4^1 = {1 \over 2 \kappa_4^2} \int d^4x e^{-2 \phi}\sqrt{g} \Biggl [
-R - 4 (\nabla \phi)^2 - {1 \over 8} {\rm Tr} \nabla M L \nabla M L
-{1\over4} F^{\hat a} (LML)_{{\hat a} {\hat b}} F^{\hat b}
+ { 1 \over 12} H^2
\Biggr ]
}
where $\phi = \Phi + \sigma$ is now the four-dimensional dilaton
field and $\kappa_4^2 = \kappa_5^2/ 2\pi R$.

For the string part of the action, note that
\eqn\gam{
{\partial X^a\over \partial \sigma^A}
{\partial X^b\over \partial \sigma^B} {\cal G}_{ab}
= h_{AB} - e^{-4 \sigma} V_A V_B
}
where
\eqn\ve{
h_{AB} = {\partial X^\mu\over \partial\sigma^A}
{\partial X^\nu\over \partial\sigma^B} g_{\mu\nu} \qquad
V_A = {\partial X^5 \over \partial \sigma^A} +
 A_A^1
}
with the pullbacks of the gauge fields given by
\eqn\pback{A_A^{\hat a} =
{\partial X^\mu\over \partial \sigma^A}A_\mu^{\hat a}.}
We also have
\eqn\bterm{
\eqalign{
\epsilon^{AB} X^a_{,A} X^b_{,B} {\cal B}_{ab} &=
\epsilon^{AB} X^\mu_{,A} X^\nu_{,B} {\cal B}_{\mu\nu}
+2 \epsilon^{AB} X^5_{,A} X^\nu_{,B} {\cal B}_{5\nu}\cr
&=-2 \epsilon^{AB} V_A A_B^2 + \epsilon^{AB}  A_A^1 A_B^2
+\epsilon^{AB}  X^\mu_{,A} X^\nu_{,B} B_{\mu\nu}
}}
So the string action reads
\eqn\strac{
\eqalign{
S_4^2 = -{1\over4\pi\alpha'} \int d^4 x d^2 \sigma
\delta^{(4)} \left ( x^\mu - X^\mu(\sigma^A) \right )
&\Bigl \{ \sqrt{-\gamma}\gamma^{AB}
[h_{AB}-e^{-4 \sigma}V_AV_B] \cr
&+\epsilon^{AB} \left [ X^\mu_{,A} X^\nu_{,B} B_{\mu\nu}
+ A_A^1 A_B^2 - 2 V_A A_B^2 \right ] \Bigr \} \cr
}}

The relevant equations of motion derived from this action
are then
\eqn\beqn{
\nabla_\mu \left ( e^{ - 2 \phi} H^{\mu\nu\lambda} \right ) =
- {\kappa^2_4 \over \pi\alpha'} \int d^2 \sigma
\delta^{(4)}(x^\mu-X^\mu(\sigma,\tau))
\epsilon^{AB} X^\nu_{,A} X^\lambda_{,B}
}
\eqn\aeqn{
\eqalign{
\nabla_\mu \left ( e^{-2 \phi} LMLF^{\mu\nu} \right )^{\hat a}
= &\quad  j^{{\hat a}\nu}\qquad\qquad\qquad \cr
= {e^{-2 \phi}\over 2 } H^{\nu\mu\lambda} (LF_{\mu \lambda})^{\hat a}
&- {\kappa^2_4 \over \pi\alpha'}\int d^2 \sigma
\delta^{(4)}(x^\mu-X^\mu(\sigma,\tau)) \left [
{T^{\hat a}}^{AB}V_A X^\nu_{,B}  \right ]
}}
where
\eqn\tsdef{T^{{\hat a}AB} = (e^{-4 \sigma}\gamma^{AB}, \epsilon^{AB})}
Note that the electric current $j^{\hat a}$ contains
two terms, the first term depending on the spacetime fields
while the second is localized on the string world-sheet.
For a string wrapped exclusively around the fifth dimension,
appearing as a particle from the four-dimensional point of view,
the current associated to $A^2$ takes a canonical  form:
\eqn\chpp{
j^{2\mu} = {\kappa_5^2\over\pi\alpha'} {dX^\mu\over dt} \delta^{(3)}
({\vec x}-{\vec X}(\tau))
}
i.e., the current appropriate to a charged particle (charge
proportional to ${\kappa_5^2\over\pi\alpha'}$)
following a trajectory $X^\mu(\tau)$.

Now consider the unwinding trajectory \unwinda.
The \kk\ monopole gives rise to a magnetic field $F^1_{\theta \phi}$,
and hence if $H_{\mu\theta\phi}$ becomes non-zero at any point current
will be present. This is precisely what happens for the unwinding string.
The string couples to $B_{\mu \nu}$ and its unwinding motion \unwinda\
gives rise to a non-zero $H_{r\theta\phi}$.
These two effects combine
to a give a radial component of the current $j^2$, that is the
charge flows in towards the center of the monopole. Meanwhile, the string
also couples to $A^2_\mu$ and gives a $\theta$-directional current on
the worldsheet so that the total current is
\eqn\overcur{
j^{2\nu} = {\kappa_5^2 U^{3/2} \over4\pi^2\alpha' r^2} \left [
{(1 + \cos\theta) \delta(r - r_0) \over \sin\theta}
{dX^\nu \over dt} - {dX^\theta\over dt}
\Theta(r_0-r) \delta_r^\nu \right ] \delta(\theta - X^\theta(t))
}
for the unwinding configuration \unwinda.

Note how
the contribution of winding-charge from the string disappears as $\theta$
moves from $0$ to $\pi$, whereas there is a steady radial inflow of
charge via $j^r$. The total inflow is obtained by integrating over
the sphere at radius $r_0$, and gives, not surprisingly,
${\kappa_5^2\over\pi\alpha'}$.
This means
that the \kk\ monopole must be able to carry the electric charge which
couples to $A^2$.

We can also understand this directly as arising from the excitation
of a collective coordinate of the \kk\ monopole. As usual, the
bosonic collective coordinates of a magnetic monopole of charge one
arise either from translations or from gauge transformations which
do not vanish at infinity. The \kk\ monopole has 3 collective
coordinates from translations in \real$^3$ but does not have a collective
coordinate from translation of $x^5$ because this is an exact symmetry
of the solution. Instead the remaining collective coordinate arises
from a gauge transformation of the antisymmetric tensor field of
the form
$\delta {\cal B} = \alpha(t) d \Lambda $
with $d{\cal B} = 0$ for constant $\alpha$ \refs{\sennew, \imamura}.
Here $\Lambda$ is a
one-form
non-vanishing at infinity such that $d \Lambda$ is a harmonic
two form -- thus guaranteeing that the deformation has zero energy
for constant $\alpha$. Explicitly we have
\eqn\deform{ \Lambda =   2 \pi R U (dx^5 + {\half} R(1-\cos \theta) d
\phi)}

We can derive an effective coupling between the string and the
\kk\ monopole collective coordinate by substituting the above
expression for ${\cal B}$ into the action \action.
This leads to the coupling
\eqn\effcou{
- {R \over \alpha'} \int d^2\sigma \alpha(t) \epsilon^{AB} \left \{
U F^1_{\theta\phi} X^\theta_{,A} X^\phi_{,B} + U' X^r_{,A} V_B
\right \} .
}
We can now see how the string excites the zero mode. For unwinding
motions at constant $r$ such as \unwinda, the $\theta$ and $\phi$
variations of the string couple via the magnetic field of the
\kk\ monopole to $\alpha$. For radial infall of the string which
remains wrapped around the $S^1$, the `particle' couples to
$\alpha$ via the geometry. In each case, suppose $X^\phi$ and $X^5$
are given by \unwinda\ (but $X^\theta, X^r$ are
arbitrary), then we
obtain the following Lagrangian for $\alpha$:
\eqn\alphlag{
{\cal L}_\alpha = {\half} {\dot \alpha}^2
+ {\alpha\kappa_4^2 \over 4\pi^2\alpha' R}\left [
U\sin\theta {d\theta\over dt} - U' (1+\cos\theta){dr\over dt} \right ]
}
For a string sitting initially at $\theta=0$ asymptotically far from
the \kk\ monopole, with ${\dot\alpha}$ consequently zero at the start
of the motion, we see that the solution for ${\dot\alpha}$ is
therefore
\eqn\dotalp{
{\dot\alpha}(t) = {\kappa_4^2\over4\pi^2\alpha' R}
\left [ 2-U(1+\cos\theta) \right ]
}
Thus, either for the string which radially infalls, or for the string
unwinding at infinity, the net result is that ${\dot\alpha}$
has increased by ${\kappa_4^2\over2\pi^2\alpha' R}$
once the string has disappeared.

\newsec{The T-dual process}

We can also consider a T-dual version of this process by performing
a T-duality transformation along the $x^5$ direction  thus working
in a background where $x^5$ has periodicity
$\tilde R = \alpha'/R$. This transformation
maps the \kk\ monopole into an H-monopole \hmon\ and a string with winding
around $x^5$ into a string with momentum along $x^5$ \tref.
The T-duality between the \kk\ and H-monopole solutions will be
discussed in more detail in the following section.

The H-monopole solution takes the five-dimensional form
\eqn\hmonsol{\eqalign{{\cal H}_{\alpha\beta\gamma} & =
2{\epsilon_{\alpha\beta\gamma}}^\delta \nabla_\delta \Phi \cr
{\cal G}_{\alpha\beta} & = e^{2 \Phi} \delta_{\alpha\beta} \cr }}
where $\alpha, \beta, \cdots$ now run over $1,2,3,5$ and $\Phi$
is given by
\eqn\phgiv{e^{2 \Phi}(x^5, \vec x) = e^{2 \Phi_0} + {\alpha' \over 2 \tilde R
r}
{ \sinh(r/\tilde R) \over \cosh(r/\tilde R) - \cos{x^5 - x^5_0 \over \tilde R}}
}
where $r = |\vec x|$ and $x^5_0$ is the location of the H-monopole
along the $S^1$.

Note in particular that the H-monopole is {\it not} invariant under
$x^5$
translations, unlike the \kk\ monopole. As a result its bosonic
collective coordinates are given by translations in \real$^3$ as well
as translation along the  $S^1$. This implies that there is a zero
mode carrying \kk\ electric charge, as expected from T-duality.

Since translation in $x^5$ is no longer a symmetry, there is no
conservation law which forbids the decay of a string with momentum
along $x^5$. However at large distance from the monopole this symmetry
is approximately valid and thus we might expect a string with momentum
at large distances from the monopole to be stable.
We can, however,
by analogy to the previous discussion write down a string trajectory
which does remove the momentum of the string while keeping it infinitely far
from from the monopole. The T-dual of the previous string trajectory
is
\eqn\unwindt{\eqalign{
& x^5= 2 \pi (\alpha'/ \tilde R) \tau  \cr
&\theta = \pi \tau  \cr
&\phi = - 2 \pi \sigma \cr
&r = r_0 \cr
 }}

To an observer looking at the string near the north pole it
 appears that this configuration carries one unit of momentum in
the $x^5$ direction. However the full expression for the momentum
$p_5$ in the fifth direction is
%
\eqn\hpfiv{p_5 = \int d \sigma  \bigl[
g_{5a} {\dot X}^a + B_{5a}{X'}^a
\bigr] }
Far from the monopole the metric is approximately flat, but we
must remember that it is necessary to use two different patches
to describe the $B$ field which is the vector potential of a Dirac
monopole. Thus to evaluate $p_5$ in the Southern hemisphere we
must use the appropriate form of $B$ which leads to
\eqn\knpf{p_5 = 2 \pi (\alpha'/{\tilde R})(1 + \cos \theta) }
which does indeed vanish at the south pole.

One way to think of this is as follows. In string theory we should distinguish
between left and right-moving coordinates on $S^1$. In the \kk\ monopole
background a string wound around
the geometrical coordinate $x^5_-= x^5_L - x^5_R $
can unwind because the $S^1$ is
fibered non-trivially over the $S^2$ at infinity.
The T-dual version has a string with
momentum along the geometrical coordinate
$x^5_+$ which is equivalent to a
string with winding around the dual
coordinate $x^5_-=x^5_L -  x^5_R$. The patching of the
$B$ field in the H-monopole is equivalent to saying
that the global topology at infinity has $S^1$ with coordinate $x^5_-$ fibered
over the $S^2$ at infinity. This illustrates
the fact that
{\it both} string coordinates
$x^5_\pm$ play a non-trivial role in the full solution.

An analysis similar to the previous one shows that \kk\ electric
charge is conserved in this process by an inflow of current
from the string into the H-monopole and excitation of the zero
mode carrying \kk\ electric charge.

\newsec{T-duality of H-monopoles  and \kk\ monopoles}

T-duality of string theory is an exact discrete gauge symmetry
of nonperturbative string theory.  We may regard both the
H-monopole and the \kk\  monopole as particular excitations
of the type II string vacua
\real$^4 \times S^1_{\tilde R} \times T^5$
and
\real$^4 \times S^1_R  \times T^5$,
respectively.
Regarded as states in string field theory these
two solutions must be identical.
There are, however, a number of
puzzling points about the relation between the two solutions as given here.

Let us call the geometrical coordinate of
$S^1_{\tilde R}$ $x^5_+$  and that of $S^1_R$ in
the T-dual background $x^5_-$.
The first puzzle is that the \kk\ monopole appears
to have an isometry while the H-monopole does not:
The H-monopole solution is
known to correspond to an exact conformal field
theory and is thus an exact solution to string theory without any $\alpha'$
corrections. As noted in the previous section it is not invariant under
translations in $x^5_+$. Thus it is not clear how to apply the standard
rules for T-duality transformations in $x^5_+$ since these require the
existence of a Killing vector field. On the other hand, the T-dual \kk\
monopole does have a Killing vector given by translation in $x^5_-$ and
is also supposed to be an exact solution of string theory since the metric
is hyperkahler. The second puzzle is that the
 H-monopole exhibits a ``throat'' metric at small values of
$\vec x$
and $x^5_+$. This is
 a reflection of the standard throat metric transverse
to a solitonic fivebrane. On the other hand the \kk\ monopole metric has
no sign of such behavior. How can these facts be reconciled with T-duality?

In this section we will argue that these puzzles are resolved by modifications
to the \kk\ monopole solution in string theory due to massive
string winding modes.
In particular, we will argue that the \kk\ monopole has a sort of throat
behavior which would be probed by scattering of string winding modes
in the same way that the H-monopole throat would be probed by the scattering
of string momentum states.

There are several ways to understand why these modes are important in the \kk\
monopole background. First, at the core of the monopole the radius of $x^5_-$
approaches zero. Thus near the the core of the monopole the string winding
modes become arbitrarily light and must be included in the dynamics.
Second, from the $T$-dual point of view,
we can Fourier decompose
the H-monopole solution by writing
\eqn\fdeom{
e^{2 \Phi}(x^5_+, \vec x) = \sum_{n=- \infty}^{+ \infty}e^{i n x^5_+/{\tilde
R}}
\Psi_n(\vec x)  }
with
\eqn\zdecom{\Psi_0 = e^{2 \phi_0} + {\alpha'\over2r{\tilde R} }}
and
\eqn\nzdecom{ \Psi_n = {\alpha' \over 2 r {\tilde R}}
e^{-|n|r/{\tilde R}}e^{-in x^5_{+,0}/{\tilde R}} }
for $n \ne 0$. The \kk\ monopole solution arises from applying
the standard T-duality
transformation laws to the metric and ${\cal B}$ field
determined from $\Psi_0$. The $\Psi_n$
can be thought of as the classical values of the \kk\  momentum states
in the H-monopole background. Note that they transform by a phase
under a shift of $x^5_{+,0}$ which is the zero mode collective coordinate for
translation in the $x^5_+$ position of the
5-brane.
Under T-duality, we would expect
that there would be classical values
for the string winding fields in the dual background. Thus under
T-duality we expect that in a background with $x^5_-$ having radius
$R = \alpha'/{\tilde R}$ and the $x^5_+$ independent fields being those
of the \kk\  monopole, that there should be classical values for
the fields for a string of winding $n$ given by
\eqn\dzma{\tilde \Psi_n = {R \over 2 r}
e^{-\vert n\vert  r R/\alpha'} e^{-i n  x^5_{+,0}
R/\alpha'} }
where $x^5_{+,0}$ is the  collective coordinate
for translations in the coordinate $x^5_+$.

The values \dzma\ of the classical winding fields
can be understood, at least qualitatively, in three
ways. Two ways are string-theoretic, the third
involves an effective field theory. The first
way involves string field theory. We regard
the \kk\ monopole as a particular state in
the type II string theory built on the vacuum
\real$^4 \times S^1_R  \times T^5$.
As such it must be  described by a vector
in the space of states $\CH_0$ of
the conformal field theory associated to this
vacuum. Thus, the \kk\ monopole must
be described by a
string wavefunction
$\Psi_{KK}[X(\sigma)]
\in \CH_0$. Concretely, the
values of the wavefunction are defined
by evaluating the worldsheet path integral
on the disk using the \kk\ background
$\sigma$-model and imposing boundary
conditions $X(\sigma, r=1) = X(\sigma)$.
On the other hand, since it is a state in $\CH_0$
we can expand in a  basis of states
$\{ \CO \} $ for $\CH_0$:
$\Psi_{KK}[X(\sigma)]
= \sum_{\CO} c_{\CO} \Psi_{\CO}$.
Moreover, we can choose an orthonormal basis with respect to the
natural bilinear pairing defined by the 2-point function
on the sphere. The values of
the coefficient $c_{\CO}$ correspond to the
values of the spacetime fields associated to
the vertex operator $\CO$. In particular,
the value of the
coefficient $c_{\CO}$ for vertex operators
creating winding strings in
\real$^4 \times S^1_R  \times T^5$ is given
by sewing into the disk amplitude defining
$\Psi_{KK}[X(\sigma)]$ the disk amplitude with
$\CO$ at its center.  In the semiclassical theory
in $\alpha'$ the effect of the vertex operator
is to impose winding boundary conditions on
the \kk\ path integral. The classical action then
gives the worldsheet disk instanton  factor
in  \dzma.

A second way to understand \dzma\ (pointed out
by E. Witten) is that in the standard argument for
$T$-duality
\tref\ref\verlinde{M. Rocek and E. Verlinde,
``Duality, Quotients, and Currents,''  hep-th/9110053,
Nucl. Phys. B373 (1992) 630-646.}\
obtained by gauging an isometry with a
$U(1)$ gauge field $A$ and
then integrating out the gauge field, instantons in
the gauge field $A$ will spoil the translation symmetry
which ordinarily forbids classical field values for
$\tilde \Psi_n$.

A third way to understand \dzma\ uses a low energy
effective field theory description.
For simplicity consider the following reduction of the problem.
We will work in a regime where we can represent the dyon degree of
freedom of the monopole as being localized at the origin. Now of all
the many fields describing excitations of the string (as in string field
theory) we keep only the fields $\tilde \Psi_n$ which create strings
with winding $n$ around $x^5_-$ far from the
monopole where, at least
locally, the metric makes spacetime look
like \real$^3 \times S^1$. We now want to
write an effective Lagrangian describing these winding fields and their
coupling to the dyon degree of freedom, $\alpha(t)$.

As a first step note that for a
{\it classical} winding mode, $x^5_- = 2\pi nR$, the effective action
can be seen to be (c.f.\ \effcou\ and \alphlag)
\eqn\classwind{
{(2\pi R)^3\over 2\kappa_4^2} \int {\half} {\dot\alpha}^2dt
- {\vert n\vert R\over\alpha'} \int \left (
{\half} e^{-2\sigma} |{\dot X}^\mu| + 2\pi R \alpha
U' {\dot X}^r \right )d\tau
}
The second part of this is the action for a charged relativistic
particle of mass $m_n = \vert n\vert R/\alpha'$,
charge proportional to $n\alpha$, in
the conformal frame ${\hat g}_{\mu\nu} = e^{-4\sigma} g_{\mu\nu}$.
Therefore, when describing the winding mode as a quantum
field, we must remember to include the correct conformal factor in
front of the mass term, as well as including the additional factor of
$e^{-2\phi} = e^{-2\sigma}$ resulting from the dimensional reduction.
\eqn\seffdy{
\eqalign{
S_{\rm eff} = & {1\over2\kappa_4^2} \int d^3x dt \left [
(D_\mu \tilde \Psi_n)^* (D^\mu \tilde \Psi_n)
+ e^{-4\sigma} m_n^2 \tilde \Psi_n^* \Psi_n \right ] \sqrt{-g}e^{-2\sigma} \cr
& + \delta^3(x) \left[ (2\pi R)^3 (\partial_0 \alpha +A_0)^2
+ (2\pi R) b (e^{i n \alpha } \tilde \Psi_n^*
+ e^{-i n \alpha} \tilde \Psi_n ) \right]  \cr}}
Since the string winding states can decay we know that there is a
mixing between $\alpha$ and the $\tilde \Psi_n$, the form of which
is dictated by gauge invariance. The dimensionless constant $b$
represents the ambiguity in replacing the $A^2_\mu$ field by a point
source at the origin,  however, note that the magnitude of this
coupling is independent of $n$, as it should be since
the geodesic motion for string winding states is independent of
$n$ \gibrub.

We can now solve for the static configuration of the
$\tilde \Psi_n$ in a background with $A_\mu^2=0$
\eqn\pssol{
(-\nabla^2 + m_n^2)\tilde \Psi_n = 2\pi R b e^{i n \alpha} \delta^3(x) }
(where $\nabla^2$ is now the flat space laplacian) with solution
\eqn\psiback{ \tilde \Psi_n = { b R\over 2r} e^{in\alpha} e^{-m_n r}
\equiv {R\over 2r} e^{in\alpha} e^{-|n|r R/\alpha'}}

This agrees with what we obtained  by T-duality if
we identify $\alpha$ with the dual
collective coordinate
for translations in $x^5_+$:
 $\alpha = -   x^5_{+,0} R/\alpha'$,
and set $b=1$.
Of course it is not clear that the correct
T-duality transformation on massive fields does not include some additional
scaling, but the argument given here should give an accurate description of
the asymptotic value of the string winding modes.

The above results suggest that the \kk\ monopole as a solution to string theory
also
exhibits the same kind of throat behavior as the H-monopole solution.
This throat could be probed by scattering string
winding states of the full \kk\ monopole solution in the same way
that the H-monopole throat can be probed by scattering string momentum
states. Unfortunately
it is difficult to give a precise description of
the throat behavior purely in field theoretic terms.
The reason is that the
collective coordinate $\alpha$ can also be thought of as a translational
zero mode for the dual stringy coordinate $x^5_+$ and it is difficult to
find a field theoretic formalism which incorporates
both $x^5_-$ as well as  $x^5_+$ as geometrical
coordinates.

\newsec{Conclusions and Outlook}

The construction we have described is clearly quite general. The global
topology of the \kk\ and H-monopoles ensures that a string with winding
or momentum can be unwound at infinity. Since the winding number is a conserved
charge, there must therefore be an excitation of the monopole which can
carry this charge. We deduce this purely from the physics at
large distances from the monopole but can then verify that this
is in fact consistent with the exact structure of the solution.
This is in some ways reminiscent of the anomaly inflow argument
of \ch\ which also allows one to deduce zero mode structure on
various defects from an analysis of couplings at large distances from
the defect.

One possible application of this mechanism is to the study
of black hole entropy. Since the analysis involves only the
large distance physics, it clearly applies quite generally
to black holes or D-brane configuration which carry \kk\ or
H magnetic charge and provides a way to change the charge of
the black hole by moving strings around at infinity. Both charged
fundamental strings and charged black holes can carry non-zero
entropy, it would be interesting to see if this mechanism can be
used to transfer entropy from a black hole to a fundamental
string at infinity. If so this might provide a concrete way to
identify the microstate of a black hole outside the D-brane regime.

We have also resolved  - at least qualitatively -   the puzzles associated with
T-duality between
\kk\ and H-monopoles. In particular, the zero mode structure is
precisely what one would expect from T-duality.
In string theory
we should distinguish between $x^5_L$ and $x^5_R$. The H-monopole
solution is not invariant under shifts of $x^5_+= x^5_L + x^5_R$ due to the
excited \kk\ momentum states but is invariant under shifts of
$x^5_- = x^5_L - x^5_R$. The same holds for
the  T-dual \kk\ monopole with the crucial difference that
the role of the stringy coordinate $x^5_-$ and the
geometrical coordinate used to write the metric
$x^5_+$ are interchanged: In the \kk\ monopole
$x^5_-$ is the geometrical coordinate, while
$x^5_+$ is stringy.  Note that the coordinate
$x^5_+$ is associated with a topologically trivial
fibration over space, while the coordinate $x^5_-$ is
only locally defined and is a fiber coordinate in
a nontrivial $S^1$ fibration. This holds true in
{\it both} the \kk\ and H-monopole backgrounds,
as required by $T$-duality.   From this understanding
of $T$-duality we are forced to conclude that
the   classical values
of the string winding fields in the \kk\ monopole background generate a  throat
which cannot however be described purely in field theoretic terms and must be
probed by scattering
with winding strings.

Finally, we note that this sort of effect is likely to lead to stringy
modifications
to other solutions of string theory which are also related by T-duality.
The throat metric and its T-dual also appear in a number of other contexts,
for example in describing the dynamics of D1-branes in the background of
D5-branes and in the compactifications of this
much-studied system. The string corrections
to the \kk\ monopole solution described here (i.e. to the Taub-Nut metric)
are likely to have application to these systems as well.

\bigskip
\centerline{\bf Acknowledgements}\nobreak

We would like to thank David Berman, Peter Bowcock, Anton Gerasimov,
Andrei Losev, Emil Martinec,  Nikita Nekrasov, Gavin Polhemus,
Samson Shatashvili,
Steve Shenker, Erik Verlinde and Edward Witten for helpful discussions.
JH and GM would like to thank  the Aspen Center for
Physics for hospitality.
This work was  supported in part by the Royal Society (R.G.),
by NSF Grant No.~PHY 9600697 (J.H.) and  by DOE grant DE-FG02-92ER40704 (G.M.).

\listrefs

\bye